\documentclass[twocolumn,pra,unsortedaddress,superscriptaddress,floatfix,citeautoscript,nofootinbib,longbibliography]{revtex4-1}

\usepackage{latexsym,epsfig,graphicx,dcolumn,subfigure,comment,ulem}
\usepackage{amsmath,eqnarray,amssymb,amsbsy}
\usepackage{xcolor,color}
\usepackage[colorlinks,urlcolor=blue,citecolor=blue]{hyperref}

\begin{document}
\title{Vortex-antivortex physics in shell-shaped Bose-Einstein condensates}

\author{Karmela Padavi\'{c}}\thanks{kpadavi2@illinois.edu}
\affiliation{Department of Physics, University of Illinois at
Urbana-Champaign, Urbana, Illinois 61801-3080, USA}
\author{Kuei Sun}\thanks{kuei.sun@utdallas.edu}
\affiliation{Department of Physics, The University of Texas at
Dallas, Richardson, Texas 75080-3021, USA}
\author{Courtney Lannert}\thanks{clannert@smith.edu}
\affiliation{Department of Physics, Smith College, Northampton,
Massachusetts 01063, USA}
\affiliation{Department of Physics, University of Massachusetts, Amherst, Massachusetts 01003-9300, USA}
\author{Smitha Vishveshwara}\thanks{smivish@illinois.edu}
\affiliation{Department of Physics, University of Illinois at
Urbana-Champaign, Urbana, Illinois 61801-3080, USA}

\begin{abstract}
Shell-shaped hollow Bose-Einstein condensates (BECs) exhibit behavior distinct from their filled counterparts and have recently attracted attention due to their potential realization in microgravity settings. Here we study distinct features of these hollow structures stemming from vortex physics and the presence of rotation. We focus on a vortex-antivortex pair as the simplest configuration allowed by the constraints on superfluid flow imposed by the closed-surface topology. In the two-dimensional limit of an infinitesimally thin shell BEC, we characterize the long-range attraction between the vortex-antivortex pair and find the critical rotation speed that stabilizes the pair against energetically relaxing towards self-annihilation. In the three-dimensional case, we contrast the bounds on vortex stability with those in the two-dimensional limit and the filled sphere BEC, and evaluate the critical rotation speed as a function of shell thickness. We thus demonstrate that analyzing vortex stabilization provides a nondestructive means of characterizing a hollow sphere BEC and distinguishing it from its filled counterpart.
\end{abstract}

\maketitle

\section{Introduction}
From the microscopic to the astronomical, shell-shaped Bose-Einstein condensates (BECs) have made their appearance in fascinating ways. In the laboratory realm of ultracold gases, the bosonic optical-lattice setting is renowned for its ``wedding-cake" structures~\cite{Batrouni2002, DeMarco2005,Campbell2006,Barankov2007,Sun2009} consisting of Mott-insulating layers sandwiching superfluid shells, while other instances of such shells include Bose-Fermi mixtures~\cite{Molmer1998,Ospelkaus2006,Schaeybroeck2009}. In superfluid helium droplets, introduction of other molecules in the center of the droplet leads to development of superfluid solvation shells around it~\cite{Grebenev2083,Whaley01}. In the stellar realm, extremely high densities in neutron stars offer the possibility of hosting condensed phases of subatomic particles, and some observed behavior has indicated shell-shaped superfluid regions~\cite{WEBER2005193,pethick2017,khomenko_haskell_2018,verma2020rotating}. BEC shells are expected to have dramatically different features from their filled counterparts in collective mode structure~\cite{Padavic2017, Sun2018, Diniz2020}, thermodynamics~\cite{PhysRevLett.123.160403,doi:10.1119/1.5125092}, and time-of-flight properties~\cite{PhysRevA.75.013611, PhysRevLett.125.010402, Meister_2019}. 
A proposed ``bubble trap" geometry~\cite{Zobay2001} in laboratory settings would allow for the realization of BEC shells in free space and the controlled tuning from a filled sphere to a topologically distinct~\cite{Sun2018, Padavic2017} hollowed out three-dimensional (3D) structure to a two-dimensional (2D) spherical surface. While so far gravitational sag has prevented the creation and analysis of free standing hollow shells on Earth, the advent of experiments in microgravity~\cite{van_Zoest1540, PhysRevLett.123.240402} holds promise. In particular, one experiment in the recently launched Cold Atomic Laboratory (CAL) aboard the International Space Station specifically designed to realize a hollow bubble geometry~\cite{Lundblad2019,frye2019boseeinstein} has provided impetus for probing the unique topological and geometric aspects of BEC shells. 

Integral to BECs, the physics of quantized vortices~\cite{RevModPhys.81.647} calls for a study in and of itself. Emergence of quantized vortices upon system rotation has been used as confirmation of superfluidity in early BEC experiments~\cite{MatthewsE99,AboShaeer476} and superfluid helium droplet studies~\cite{Gomez906,Gessner19} alike. In shell-shaped structures in particular, tuning through the dimensional crossover from a filled sphere to a thin shell, a single vortex line passing through the filled system reduces to a vortex-antivortex pair on the 2D shell surface. The associated thermodynamics is highly sensitive to the closed geometry and topology, particularly with regards to vortex-antivortex induced Berezinskii-Kosterlitz-Thouless transition~\cite{Kosterlitz_1973}. In the presence of stirring or rotation, static and dynamic features of vortices in shell BECs are expected to be significantly altered compared to filled sphere counterparts. An extreme instance involves possible explanations for pulsar glitches attributed to tangles of vortices in neutron stars~\cite{khomenko_haskell_2018,verma2020rotating}. Here, we demonstrate that even the simplest vortex configuration in a spherical shell geometry exhibits rich physics due to interplay between restricted flow in a closed topology and the energetics of vortices in hollow structures. Topological constraints enforce a zero circulation rule, applicable even to classical fluids such as planetary atmospheres~\cite{TOZZI20201}, requiring that the vortex-antivortex pair be the minimally allowed configuration~\cite{lamb1945hydrodynamics, PhysRevD.43.1314}, in contrast to a flat 2D system where a single isolated vortex is possible.  We investigate equilibrium properties of such a pair across the finite thickness to strictly 2D shell regimes, establishing that upon rotation they are stable only when at the poles.  The energy barrier for stabilizing such a pair in a shell always remains less than for a filled sphere of the same radius, as expected from the absence of core energy in the hollow region.  Along with previous predictions for collective mode excitations in BEC shells~\cite{Padavic2017,Sun2018}, these studies provide a nondestructive means for characterizing a hollow sphere and distinguishing it from its filled counterpart. 

\begin{figure}[t]
\centering
\includegraphics[width=8.6cm]{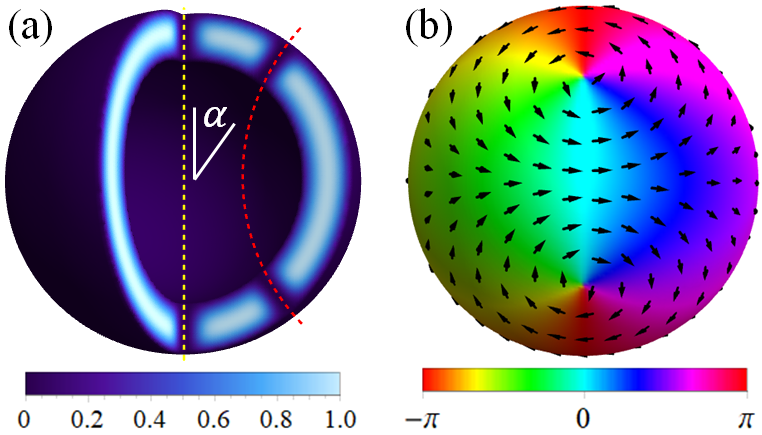}
\caption{(a) Schematic density profile of a shell-shaped BEC comparing a vortex-antivortex pair at the poles, rotating about the $z$ axis (straight dashed line) with another such pair at polar angle $\theta = \alpha$ (antivortex at $\pi - \alpha$) rotating about a curved dashed line. (b) Vector field showing superfluid flow for the vortex-antivortex pair at $\theta = \alpha$. The $+x$ direction points out of page and the colors represent the phase $S$ of the BEC wave function. }
\label{fig:1}
\end{figure}

Our starting point for analyzing vortex structures in shell-shaped BECs assumes a condensate wave function $\psi=fe^{iS}$, where $S$ represents the condensate wave function phase and $f$ its amplitude. For the 2D and 3D cases, $f$ is related to the condensate density $f^2=\rho_{\mathrm{2D}}$ or $\rho$, respectively. The condensate velocity takes the form $v=\hbar \nabla S/m$ for condensate atoms of mass $m$. 
Near the vortex center, the rotation speed of condensate atoms exceeds the Landau criterion for destruction of superfluidity and a vortex ``core" is formed. The size of this density depletion is set by the condensate's healing length $\xi_0$, which depends on the strength of interatomic interactions. 

In what follows, we first consider a 2D BEC confined to the surface of a sphere, detailing the topological constraints on and interactions between vortices in this closed, curved geometry. We find that long-range attractive interactions, induced by the energy gradient, tend to energetically drive a vortex and an antivortex towards the equator, while rotation of the system tends to drive them to opposite poles. Within a Gross-Pitaevskii (GP) formalism that allows us to include the interactions between condensate atoms and the nonzero size of the vortex cores, we find the critical rotation speed necessary to stabilize the pair. We chart the condensate flow pattern in the presence of the associated vortex line threading through the 2D surface system, as shown in Fig.~\ref{fig:1}(a). Turning to thin 3D hollow shells, we use a simple slicing argument and a local density approximation to argue that a critical rotation speed for stabilizing a single vortex line persists in this more realistic curved hollow system and show that its value increases with the thickness of the shell. We perform a fully 3D numerical solution of the equilibrium GP equation in order to obtain energetic estimates of the critical rotation speed for a hollow shell away from the thin-shell limit and compare it to the case of a filled sphere. We conclude with a brief outlook on dynamics and multivortex scenarios in the context of BEC shells.

\section{Vortices on a two-dimensional spherical surface}
While the precise distribution of vortices and superfluid flow patterns for a hollow condensate shell depends on its geometry and detailed energetics, the $S^2$ topology 
associated with the closed surface poses significant constraints~\cite{Guenther20}. By way of illustration, we first discuss an infinitesimally thin, effectively 2D shell BEC. Consider a collection of $n_v$ pointlike vortices on the surface of the sphere having integer vorticities $\ell_i$, respectively. Units of circulation  
within any closed loop on the surface can be counted using a loop integral, equivalent to a surface integral via Stokes' theorem ($\oint \nabla S\cdot dl=\oint \frac{\hbar}{m}v \cdot dl$). If the loop is chosen to contain all vortices,  then the rest of the shell contains no vortices and we identify the constraint of vanishing net vortex circulation,  $\sum_{i=1}^{n_v}\ell_i=0$. Such constraints have been identified for a broader class of compact surfaces, for instance, toroidal surfaces of revolution in Ref.~\cite{Guenther20}. The simplest instance of vortex stabilization on the surface of a spherically symmetric BEC, the focus of this work, then is $n_v=2$ and $\ell_1=-\ell_2=1$~\cite{lamb1945hydrodynamics,PhysRevD.43.1314}. This describes a vortex carrying a single unit of angular momentum in tandem with an identical vortex having the same vorticity but oriented in the opposite direction, i.e., a vortex-antivortex pair.

Further, a nonzero flow field on the condensate shell has to satisfy the Poincar\'e-Hopf theorem, which states that for a vector field everywhere tangent to the surface of a sphere, the topological indices associated with its zeros and singularities must add up to its Euler characteristic. This theorem is equally applicable to fluid flow in other settings, such as the Earth's atmosphere~\cite{TOZZI20201}. Explicitly, denoting a topological charge of the $i$th defect in a collection of $n_d$ vortices and singular flow points on the surface of a sphere as $q_i$ implies $\sum_{i=1}^{n_d}q_i=2$. Regardless of its vorticity, $\ell_i$, a point vortex here has topological charge $q_i=1$, while stagnation points in the flow contribute $q_i=-1$. The vortex-antivortex pair ($q_i=1$ each) in the absence of any stagnation points in the superfluid flow is thus the simplest allowed vortex configuration for the condensate shell under the constraints posed by its geometry and topology. 

Two pointlike vortices carrying opposite vorticities in a 2D condensate experience an attractive interaction. Intuitively, interactions between vortices stem from their being sources of vorticity for the superfluid flow---as the superfluid circulates around each vortex, this motion carries other vortices with it. In this sense, vortex interaction energy reflects the kinetic energy of the underlying condensate. Similar physics occurs for a collection of current carrying wires where each wire is affected by a Lorentz force due to magnetic fields produced by other wires~\cite{Fetter1966}. We note, however, that the dynamical behavior of two wires carrying opposite currents is different than that of a pair of vortices carrying opposite vorticities as the former repel while the latter do not. To make the superfluid vortex case more precise, as in Refs.~\cite{Vitelli2006,Turner2010}, we write this effective flow-mediated interaction energy as 
\begin{eqnarray}\label{eq:KE}
E_{\rm{v}\mbox{-}\rm{v}}=\frac{\hbar^2\rho_{\mathrm{2D}}}{2m^2}\int |\nabla S|^2dA
\end{eqnarray}
We evaluate this expression by solving for the inverse of the Laplacian
on the spherical shell, i.e., by finding the Green's function as outlined, in more detail, in Ref.~\cite{Turner2010}.

For a vortex-antivortex pair situated at $(\theta,\phi)=(\alpha, 0)$ and  $(\pi - \alpha, 0)$, respectively (see examples in Fig.~\ref{fig:1}), i.e., a dipolelike configuration symmetric about the equator of the BEC shell, this interaction energy takes the form
\begin{eqnarray}\label{eq:KEalpha}
E_{\rm{v}\mbox{-}\rm{av}}(\alpha)=\frac{\pi\hbar^2\rho_{\mathrm{2D}}}{m^2}\ln(\cos\alpha).
\end{eqnarray}
Here, the shell topology is reflected in the logarithmic scaling of interaction energy with the angular separation of vortices, rather than with the rectilinear distance between them, which is the case in flat superfluid topologies. For the vortex-antivortex pair, the condensate kinetic energy decreases as $\alpha \to \pi/2$,  resulting in an attractive interaction. In the presence of energy and angular momentum dissipation mechanisms~\cite{pethick_smith_2008}, the vortex and the antivortex will tend to relax towards $\alpha=\pi/2$, the equator. In the full $\alpha=\pi/2$ limit,  the vortices overlap and annihilate at the equator and the flow sourced by either, and its associated kinetic energy, vanishes. A vortex-antivortex pair in a flat 2D superfluid experiences a similar attraction~\cite{Calderaro17,Fetter66}.

\begin{figure}[t]
\centering
\includegraphics[width=6.5cm]{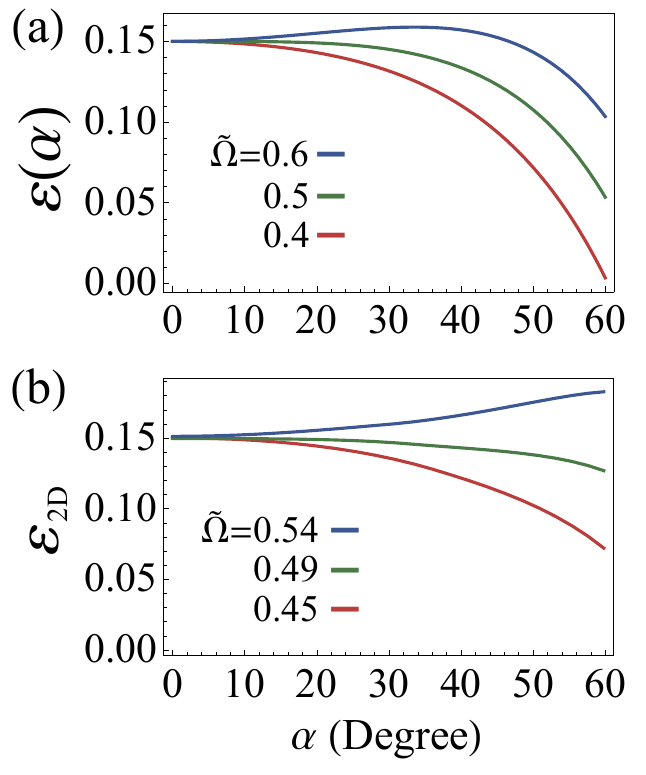}
\caption{(a) Dimensionless superfluid flow-based energy $\varepsilon (\alpha)$ [in Eq.~(\ref{eq:Efunctional})] of a rotating 2D spherical BEC for dimensionless angular velocity $\tilde \Omega$ for values that, from the top to the bottom, are (i) above, (ii) equal to, and (iii) below the critical value $\tilde \Omega_c = \frac{1}{2}$, where a local minimum develops for a polar vortex-antivortex pair. (b) Dimensionless energy ${\varepsilon_{{\rm{2D}}}}$ [in Eq.~(\ref{eq:2D_energy_functional})] obtained from the numerical GP calculations. The curves from the top to the bottom are for $\tilde \Omega = 0.54$, 0.49, and 0.45, respectively. For ease of comparison, we have shifted energy curve offsets so as to align at $\alpha = 0 $. The dimensionless energies are in units of the characteristic energy $E_R = \hbar^2/(2 m R^2)$, where $R$ is the spherical radius.}
\label{fig:2}
\end{figure}

Given this attractive interaction, we investigate whether external rotation of the system can stabilize the pair against energetically relaxing towards the equator. For a rotating system, the total energy of vortices decreases as the system's angular momentum increases. Since angular momentum is maximized when the two vortices align with the rotation axis, we expect a minimum of the total vortex energy for $\alpha=0$ and sufficiently fast shell rotation. Heuristically, transforming the shell energy to a rotating frame with angular velocity $\Omega$, $E\to E-\Omega\langle L_z\rangle$ implies the same conclusion.

To formalize this intuitive understanding that condensate flow on the 2D shell pulls vortices having opposite circulations closer together, but rotation pushes them apart and towards the rotation axis, we evaluate the energy associated with rotation to be
\begin{eqnarray}\label{eq:Erot2}
 E_{\mathrm{rot}}&&=\Omega\langle L_z\rangle=\Omega\rho_{\mathrm{2D}}\int R^2\sin\theta d\theta (d\vec{l}\cdot\vec{v})_{\phi}\nonumber\\
&& =4\pi\frac{\hbar}{m}\rho_{\mathrm{2D}}\Omega R^2\cos\alpha.
\end{eqnarray}
where $R$ is the shell radius. For $\alpha=0$, Eq.~(\ref{eq:Erot2}) reduces to the number of atoms on the surface of the sphere rotating about the antipodal vortex-antivortex pair. For $\alpha \neq 0$, the factor of $R^2\cos\alpha$ indicates a projection of angular momentum to the rotation axis of a given vortex, i.e., its center. Combining Eqs.~(\ref{eq:KEalpha}) and~(\ref{eq:Erot2}), the dimensionless form of the energy per particle in the rotating frame becomes
\begin{eqnarray}\label{eq:Efunctional}
 \varepsilon (\alpha) = \frac{{{{E}_{\rm{v}\mbox{-}\rm{av}}} - {{E}_{{\rm{rot}}}}}}{NE_R} = \frac{1}{2}\ln (\cos \alpha ) - \tilde \Omega \cos \alpha ,
\end{eqnarray}
where $N = 4 \pi R^2 \rho_{\mathrm{2D}}$ is the total number of condensed atoms, $E_R = \hbar^2/(2 m R^2)$ sets the characteristic energy scale, and $\tilde \Omega = \frac{2m}{\hbar} \Omega R^2$ is a dimensionless angular velocity. 

Based on this energetic form, Fig.~\ref{fig:2}(a) shows that the energy functional exhibits differing behaviors for $\tilde \Omega$ above or below a critical value $\tilde \Omega_c = \frac{1}{2}$. The global minimum of $\varepsilon (\alpha)$ is always at $\alpha=\frac{\pi}{2}$. If $\tilde \Omega \le \tilde \Omega_c$, $\varepsilon (\alpha) $ monotonically decreases with $\alpha$, again implying the tendency of energetically relaxing towards the equator at $\alpha=\frac{\pi}{2}$. If $\tilde \Omega >\tilde \Omega_c$, $\varepsilon (\alpha)$ develops a local minimum at $\alpha=0$ and a local maximum at $\alpha= \alpha_M= \cos^{-1}(1/2\tilde \Omega)$, decreasing for $\alpha_M < \alpha < \frac{\pi}{2}$. Any vortex-antivortex pair initially located at $0 < \alpha < \alpha_M$ tends to stabilize along the rotation axis, while for $\alpha > \alpha_M$ the pair is unstable and tends to relax towards $\alpha=\frac{\pi}{2}$ (the shell's equator) regardless of system rotation. However, for angles near $\alpha=\frac{\pi}{2}$, we may no longer disregard vortex core effects, and this energy functional no longer holds.

To complete our analysis of the 2D vortex-antivortex pair energetics, we next include the effects of actual interactions between the atoms in the BEC. This also implies that we consider vortices as having density depletion at their cores over the size of the healing length $\xi_0$ set by the strength of these interactions instead of being pointlike. We numerically calculate the wave function $\psi_{{\rm{2D}}} (\theta, \phi)$ of a rotating 2D spherical BEC by minimizing the energy functional in the rotating frame (the standard GP formalism~\cite{2D_numerical_method}),
\begin{eqnarray}\label{eq:2D_energy_functional}
\varepsilon_{\rm{2D}}[\psi_{\rm{2D}}] &=& {R^2} \int \sin \theta d\theta d\phi \big [ |\nabla \psi_{\rm{2D}}|^2 + \frac{U_{\rm{2D}}}{2}| \psi_{\rm{2D}}|^4 \nonumber\\
&& + \tilde \Omega \psi _{{\rm{2D}}}^*(i{\partial _\phi }){\psi _{{\rm{2D}}}}\big],
\end{eqnarray}
which is rendered dimensionless by $E_R$. Here, $U_{{\rm{2D}}}$ is the dimensionless 2D effective interaction strength, which is set to keep the typical ratio of kinetic energy to interaction energy to be $\sim 5\%$ (e.g., $U_{{\rm{2D}}}=1000$ in our simulation). The vortex-antivortex pair configuration is imposed by fixing the wave function zeros (vortex cores) at $\alpha$ and $\pi-\alpha$. In Fig.~\ref{fig:2}(b), we plot the energy functional obtained by this GP calculation. The curves exhibit behavior above and below a critical angular velocity $ \tilde \Omega_c \sim \frac{1}{2}$ consistent with Fig.~\ref{fig:2}(a), confirming the stability of the vortex-antivortex pair at the condensate shell poles for sufficiently fast rotation. Numerical results in Fig.~\ref{fig:2}(b) also show the energy functional's decrease toward the global minimum at $\alpha = \pi/2$ to be less drastic than suggested by the analytic results in Fig.~\ref{fig:2}(a) [Eq.~(\ref{eq:Efunctional})]. This reflects the effect of density depletion at the vortex cores moderating the intervortex interaction at close separations.

Focusing on the superfluid flow obtained from the GP wave function, for any finite $\alpha$ we find that the flow pattern resembles one that rotates about a string through both vortex cores, as illustrated in Fig.~\ref{fig:1}. We employ a variational approach to approximate the wave function by considering the shape of such a string, parametrized in a convenient way. This allows us to determine the condensate density and phase, and the associated superfluid flow. We set an arbitrary conic-section curve on the $x$-$z$ plane ($\phi=0$ or $\pi$) through the vortex cores on the sphere, which takes the general form
\begin{eqnarray}
{z^2} = (\lambda  - 1){x^2} + 2bx + ( 1 - \lambda {\sin ^2}\alpha  - 2b\sin \alpha ),
\end{eqnarray}
with two variational parameters $b$ and $\lambda$. 
We numerically minimize the energy functional of Eq.~(\ref{eq:2D_energy_functional}) with a variational wave function $\psi_{\rm{var}} (\theta, \phi) = f_{\mathrm{var}}e^{iS_{\mathrm{var}}}$, where the phase $S_{\mathrm{var}}$ is now the azimuthal angle with respect to the variational curve and the amplitude $f_{\mathrm{var}}$ has a variational depletion at the vortex cores. Specifically, 
\begin{eqnarray}
f_{\mathrm{var}}(\theta, \phi) = A\frac{{{\sigma _\alpha }}}{{\sqrt {{\zeta ^2} + {\sigma _\alpha }^2} }} \times \frac{{{\sigma _{\pi  - \alpha }}}}{{\sqrt {{\zeta ^2} + {\sigma _{\pi  - \alpha }}^2} }},
\end{eqnarray}
where $A$ is the normalization constant, $\zeta$ a variational parameter, and ${\sigma _\alpha } = \cos^{-1} (\cos \alpha \cos \theta  + \sin \alpha \sin \theta \cos \phi )$ is the distance from the vortex core at $\alpha$. We find that the minimum energy state corresponds to a circle string ($\lambda = 0 $) passing through the vortex cores and perpendicular to the sphere's surface. Further, these variational results qualitatively agree with the GP results with respect to energy functional curves as well as critical rotation speed. 

The variational assumption of a curved string connecting the vortex-antivortex pair relates our analysis to studies of vortex dynamics in filled 3D spherical BECs where a physical vortex string (line) may move off axis~\cite{pethick_smith_2008,Rosenbusch02} or bend~\cite{Aftalion01,Aftalion03,Modugno2003} due to interaction or density inhomogeneity effects. Namely, a vortex-antivortex pair on opposite poles of a 2D condensate shell corresponds to a straight on-axis vortex in a 3D BEC, while the pair at angle $\alpha > 0$ corresponds to a bent vortex. This correspondence additionally informs the understanding of vortices on a spherically symmetric BEC undergoing dimensional crossover from a hollow 2D geometry to a shell-shaped condensate having finite thickness and, finally, to a fully filled spherical BEC. Such crossover could be experimentally achieved by the bubble trap~\cite{Zobay2001} which we model in the next sections of this work.

The 2D condensate shell, as we have shown so far, captures rich physics in and of itself. Here, flow-based vortex-vortex interactions cause the simplest vortex arrangement allowed by the shell-shape---the vortex-antivortex pair---to purely attract in the absence of rotation. Vortices can be stabilized against the attraction by rotating the system above some critical rotation speed and, with the view of dimensional crossover, the same stabilization would occur for an equivalent single vortex line along the rotation axis in a condensate shell away from the truly 2D limit. Interatomic interactions within the condensate, and consequent nonzero vortex core size, further lessen the attractive interaction between the vortex-antivortex pair at short intervortex separations.  

\section{Thin three-dimensional shells} \label{sec:3d_shell}

Expanding on our observations in the 2D limit, we analyze the realistic case of a thin but not perfectly 2D spherical condensate shell containing a vortex line (the natural extension of the 2D shell vortex-antivortex pair) from multiple perspectives. First, using a Thomas-Fermi (TF) approximation, we compare the energetics in the thin shell with that in the case of a filled shell. We next approach the thin shell from the 2D limit by building up the former as several layers of latter. Finally, we use the 3D GP energy functional analogous to Eq.~(\ref{eq:2D_energy_functional})~\cite{pethick_smith_2008} and include a bubble trapping potential to rigorously corroborate the comparison and obtain accurate estimates for critical rotation frequencies. 

To estimate the energy cost of a vortex line in a shell-shaped condensate and compare with a filled sphere, we closely follow the approach for the latter in Ref.~\cite{pethick_smith_2008}, 
which is expected to be valid in the TF limit, i.e., when the condensate healing length is small compared to its radius (or thickness).  The situation considered here, a single vortex line along the $z$ axis having winding $\ell$, can be described by the BEC wave function $\psi({\bf{r}})=f(r_{\perp},\theta)e^{i\ell \phi}$ (in cylindrical polar coordinates). Assuming an $\ell=1$ vortex, we imagine slicing the condensate (either filled sphere or hollow shell) into thin sections of height $dz$, then integrating over $z$. For a harmonically confined system (filled-sphere BEC), where each slice is a disk BEC pierced through its center by the vortex core, Ref.~\cite{pethick_smith_2008} finds the fractional energy cost of the vortex to be
\begin{equation}\label{eq:slicing_sphere}
E_{\rm{v}}^{\rm{sphere}}/E_0^{\rm{sphere}} \approx \frac{4\pi \hbar^2}{3mU}R \left(\ln{\frac{R}{\xi_0}} -0.399 \right).
\end{equation}
Here, $U$ is the 3D interatomic interaction strength, $\xi_0$ is the coherence length at the center of the vortex-free BEC, and $R$ is its outer radius. We note that the vortex core size is set by $\xi_0$, which should be much smaller than the size of the cloud, $R/\xi_0 \gg 1$.

For a thin spherical shell having thickness $\delta$, much smaller than outer radius $R$, a similar slicing procedure now includes two disks (at the top and bottom of the system) and a number of intervening thin annuli, all with the vortex threading through their centers, perpendicular to the slicing plane. The resulting fractional energy cost of the vortex in this system is:
\begin{equation}\label{eq:slicing_shell}
E_{\rm{v}}^{\rm{shell}}/E_0^{\rm{shell}} \approx \frac{2\pi \hbar^2}{3mU}\delta \left(\ln{\frac{R}{\xi_0}}+\ln{\frac{\delta}{\xi_0}}+4.597\right),
\end{equation}
where $\xi_0$ is the coherence length at the mean radius of the shell and we note that, as above for the sphere, we assume $\delta/\xi_0 \gg 1$. We note that since $R \gg \delta$ for a thin shell, the first term in parentheses dominates.

Based on this analytical calculation we conclude that the energy cost of a vortex in a thin shell BEC scales linearly with its thickness. In contrast, the energy cost of a vortex in a filled sphere BEC scales linearly (with log corrections) with its radius $R$. This result suggests that the dominant cost of a vortex is its core, which has length $2R$ in the filled sphere and only $2\delta$ for the thin shell. Since in the thin shell limit we assume a hollow BEC thickness much smaller than its radius, the energy cost for a vortex in this geometry will be much lower than for a similarly sized fully filled spherical condensate.

\begin{figure}[t]
\centering
\includegraphics[width=6.5cm]{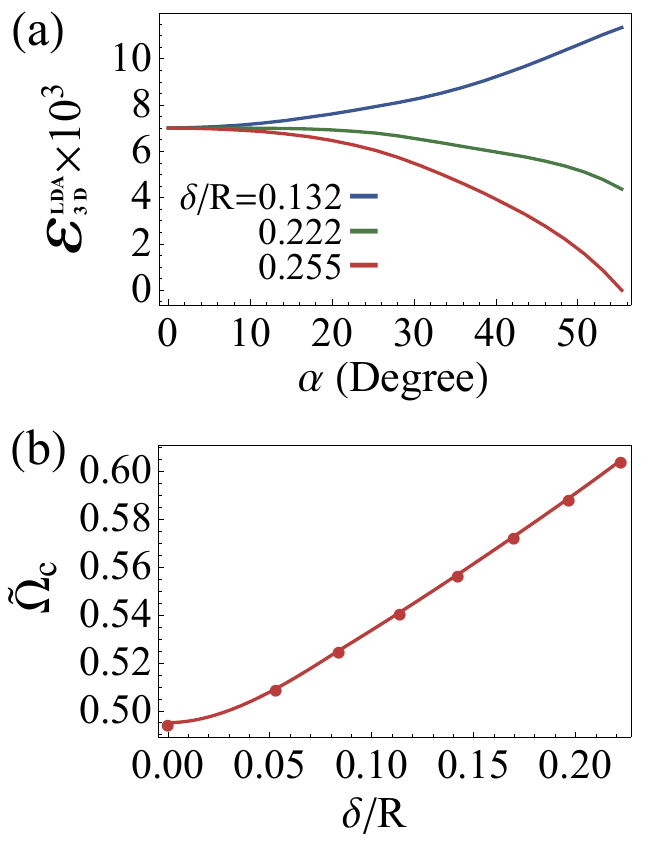}
\caption{(a) Dimensionless energy $\varepsilon_{{\rm{3D}}}^{{\rm{LDA}}}$ [in Eq.~(\ref{eq:LDA_energy})] of 3D shell BECs as a function of vortex location for various shell thickness $\delta$ and dimensionless rotation speed $\tilde \Omega = 0.6$. The curves from the top to the bottom are for $\delta/R = 0.132$, 0.222, and 0.255, respectively. The data are from the LDA calculation and are off set to level up at the $\alpha = 0 $ point. (b) Critical rotation speed $\tilde \Omega_c$ versus shell thickness from the LDA calculation. The leftmost data point is for a pure 2D shell.}
\label{fig:3}
\end{figure}

Having compared the hollow shell with the filled sphere, we now turn to the opposite limit of using a local density approximation (LDA) to decompose a thin 3D shell into layers of concentric 2D shells with radii ranging between the 3D BEC's inner Thomas-Fermi radius $R_{\rm{in}} = R - \delta$ and its outer Thomas-Fermi radius $R$. Within this approach, a vortex line in the 3D condensate shell (spanning its thickness) is equivalent to a stack of 2D vortices, each associated with one concentric shell layer. The system's energy functional is evaluated as
\begin{equation}\label{eq:LDA_energy}
\varepsilon_{{\rm{3D}}}^{{\rm{LDA}}} = \sum\limits_{i} {\varepsilon_{{\rm{2D}}}^{{\rm{}}}({r_i})}.
\end{equation}
Here, ${\varepsilon_{{\rm{2D}}}^{{\rm{}}}({r_i})}$ is the energy of each 2D layer, taking the same form as Eq.~(\ref{eq:2D_energy_functional}) except with the layer radius equal to $r_i$. The effective rotation speed for each 2D layer is scaled here by its radius as ${{\tilde \Omega }^{{\mathop{\rm LDA}\nolimits} }}({r_i}) = (r_i^2/{R^2})\tilde \Omega$, thus contributing an $r$-dependent rotational energy to the total energy functional. Therefore, there ought to exist a radius $r_c$, such that the effective rotation speed is large enough to stabilize a vortex pair on the poles for the layers with $r_i>r_c$ (the ``outer'' layers) but not for those with $r_i<r_c$ (the ``inner'' layers).  Since the vortex line cannot break into parts, its stability is determined by the layers that energetically dominate and thus depends on both the rotation speed $\tilde \Omega$ (which determines $r_c$) and the shell thickness $\delta$ (which determines the relative energetic contributions between the outer and inner layers).

Given a fixed rotation speed, the thicker the shell, the more the inner layers contribute to the system's energy. Thus, we expect to identify a critical thickness beyond which the local energy minimum for a vortex line piercing through the poles disappears.  Figure \ref{fig:3}(a) shows the energy functional of shell BECs having various thicknesses at $\tilde \Omega = 0.6$ (with the vortex location fixed at $\theta = \alpha$ on each 2D layer). The critical thickness above which the vortex line is no longer stable at the poles in this case corresponds to $\delta/R=0.222$. 

Conversely, for a given thickness, the LDA calculations show a critical rotation speed that determines the stability of the vortex line through the poles. In Fig.~\ref{fig:3}(b), we plot the critical rotation speed as a function of BEC shell thickness, and find an approximately linear relationship. This result suggests a nondestructive way to experimentally probe the thickness of a BEC shell by finding the lowest rotation speed that stabilizes a single vortex aligned with the rotation axis and passing through the poles.

In using the LDA above, we ignore any possible coupling between the 2D layers. This includes inter-layer particle movements that would result in a radial superfluid current and associated kinetic energy. For a rotating 3D shell BEC having thickness similar to the critical value (at that rotation rate), we expect a very flat energy functional indicating an approximate tie in the competition between outer and inner layers. In such a case, the radial energy, mostly coming from the tilting or bending of the vortex line, might play a significant role in vortex stability. However, for thin enough shells, we do not expect these effects to be important; for thin condensate shells, effects beyond the LDA should not qualitatively change the vortex behavior found in the LDA calculations.

To generalize our discussion further and address the limitations of not only the Thomas-Fermi approximation but also the LDA, we turn to a numerical solution of the 3D GP equation with a vortex having vorticity $\ell$ along the $z$ axis. The resulting energy equation for the condensate wave function magnitude, $f$, can be minimized using an imaginary-time algorithm~\cite{Chiofalo}. By taking the confining potential to be a bubble trap, we can numerically solve for the ground-state wave function amplitude and energy for a harmonically trapped filled spherical condensate, as well as a hollow BEC shell. The choice of the bubble trap, in particular, makes this calculation relevant for a large class of spherically symmetric condensates of arbitrary size, hollowness, and thickness. 

In order to compare with our calculations above, we report results on a fairly thin shell with thickness-to-radius ratio $\delta/R \approx 0.3$ and confinement frequency $\omega_0$. We compare these thin-shell results to the case of a filled sphere BEC trapped with the same confinement frequency. In order to compare with the Thomas-Fermi results presented thus far, we work with a relatively large interaction strength: for the spherical and shell systems described above, without a vortex imposed, the ratio of kinetic to interaction energy in the ground state varies through the range 2\%--20\% (from the filled sphere to the thin shell). Further details can be found in the Appendix \ref{app:numerics}.

By numerically obtaining the ground-state energy for the no-vortex ($\ell=0$) and single-vortex ($\ell=1$) cases, we obtain the critical rotation for stabilizing a vortex---at this rotation rate, the energy of the no-vortex state and the single-vortex state are equal. For the thin ($\delta/R \approx 0.3$) shell, we find $\Omega_c = 0.02 \omega_0$, whereas for the filled sphere we find $\Omega_c = 0.2 \omega_0$. This factor of ten difference in critical rotation speeds bears out and illustrates the much lower energy cost for a vortex in a hollow shell, compared with a filled sphere, in accordance with our previous discussions. Recent experiments aboard the Cold Atom Lab~\cite{Lundblad2019} working to create a hollow shell-shaped condensate have estimated confinement frequencies of 100--1000 Hz, giving a predicted critical rotation speed of 2--20 Hz from both the simple slicing argument in Eq.~(\ref{eq:slicing_shell}) and the imaginary time numerical calculations for the BEC shell.

To consolidate our findings for thin shells and their implications, in considering a straight vortex line connecting the poles of a spherically symmetric condensate, we always expect a much smaller energy cost for this vortex configuration in a hollow BEC than in a filled one based on comparing the length of the vortex core in the two systems. We have shown this to be the case through a simple slicing argument and a more general numerical method. We therefore expect that vortices will be much more energetically favorable in a hollow shell system than in a filled system of the same outer shape. Consequently, as with flat nearly 2D BEC layers compared to more 3D bulk structures, the spontaneous appearance of vortices in a hollow system should happen at a lower temperature than for a filled system. This is a reflection of the Berezinskii-Kosterlitz-Thouless superfluid transition temperature~\cite{Kosterlitz_1973} being lower than the BEC transition temperature in two dimensions.

Using a LDA and a fully 3D numerical method, we have further shown that there is a critical rotation speed necessary to stabilize a vortex line along the rotation axis and that this critical rotation speed is much smaller for a hollow shell BEC than for a filled-sphere condensate of the same number of particles. Since a vortex-antivortex pair on opposite poles of a 2D condensate shell has a 3D counterpart in the vortex line extending along the rotation axis, this result highlights the validity of our reasoning across hollow BECs of differing dimensionality. Finally, we note that, as the LDA results show that a local minimum develops at $\alpha = 0$ (and at no nonzero angle) at the critical rotation frequency, the hollowness of a system could be probed experimentally by looking for vortex stabilization as a function of rotation or stirring speed. 

\section{Outlook and Future Work}

While our focus has been on vortex stabilization and equilibrium features, dynamic considerations may be crucial and complex, as found in fully filled condensates. In these cases, off-axis  vortex lines are unstable even in spherical geometries in the presence of dissipation which serves to move the vortex line from a local to a global minimum~\cite{pethick_smith_2008}. They tend to precess in radially symmetric 2D geometries~\cite{Kevrekidis17} and bend in cigar-shaped condensates~\cite{Aftalion01,Aftalion03,Modugno2003}. We note that the ``slicing and stacking" method used above, combined with these dynamical results, would imply that a vortex line nucleated at a lateral off-set from the rotation axis in a thin condensate shell would be unstable to bending or dissipation-driven motion towards the outer edge of the condensate~\cite{Fetter1967}. This can be validated using the method of images to show that a point vortex in a 2D annular BEC has a nonzero velocity depending on condensate radii~\cite{Fetter1967,Guenther17}. In highly dissipative shell-shaped condensates, one could characterize the dissipation-driven motion of the vortices with the energy functional in Eq.~(\ref{eq:Efunctional}) or those computed numerically  (as shown in Figs.~\ref{fig:2} and \ref{fig:3}). A trajectory $\alpha (t)$ that the vortices follow on the sphere could be mapped out from the energy functional $E(\alpha)$ and a given energy dissipation with time $E(t)$. This approach would be complementary to existing literature employing the dissipative Gross-Pitaevskii equation for studies of, for example, vortex-driven superfluid turbulence~\cite{Billam15}. For a less dissipative system, dynamical behavior of the vortex-antivortex pair, possibly moving in concert as a dipole having a fixed cord length~\cite{Newton2001TheNP}, would be of particular interest as well. In the regime of high condensate rotation speed could stabilize a vortex line against bending or excitations (for instance,  Kelvin waves~\cite{Svidzinsky00,Cooper08}) through angular momentum effects similar to those discussed above. Various experiments in related contexts hint at this behavior, such as in Ref.~\cite{Rosenbusch02}, where a vortex line in a prolate harmonically trapped BEC becomes more bent  and deviates more from the center of the condensate as the angular momentum of the system decreases.

Furthermore, many BEC situations of physical relevance are likely to involve complex multivortex dynamics. Vortices are often nucleated as a consequence of perturbing condensates; the dynamical instability of vortex lines presents a starting point for understanding subsequent reequilibration processes. In systems with fast rotation, equilibrium and dynamic behavior induces a range of structures, including vortex lattices~\cite{Feder99,Cai18,Aftalion05}, giant vortices~\cite{Kasamatsu02}, and tangles of vortex lines. In the context of BEC shells too, we expect diverse dynamic and multivortex phenomena that are modified in comparison with their filled counterparts due to the different geometry and topology. For instance, compared to the vortex lattices observed in pancake-shaped or filled spherical condensates, the nature of the lattices in hollow shells would be modified by the differences stemming from topology, curvature, dimensionality, and shell thickness. In principle, these phenomena could be systematically investigated in the CAL experiment in the future, starting with the vortex-antivortex situation and scaling up to several vortices. Our study also offers a first step towards deconstructing theoretical and experimental multivortex dynamic studies in a range of settings from CAL and other ultracold atomic systems to pulsar glitches observed in the context of neutron stars~\cite{khomenko_haskell_2018,verma2020rotating}, where Gross-Pitaevskii numerical schemes have already proven to be relevant~\cite{Warszawski11}.

\acknowledgements
We thank Nathan Lundblad for illuminating discussions. KP, SV, and CL acknowledge support by NASA (SUB JPL 1553869 and 1553885). SV thanks UC San Diego for its hospitality through the Burbidge Visiting Professorship program.  
\appendix
\section{3D GROSS-PITAEVSKII NUMERICS}\label{app:numerics}

Here we provide additional detail on the numerical solution of the 3D Gross-Pitaevskii (GP) equation whose results are presented in Sec.~\ref{sec:3d_shell}. We describe the equilibrium (ground state) condensate wave function $\psi ({\bf{r}})$ with the standard GP equation, given by
\begin{eqnarray}\label{eq:3DGPeqn}
\left[ { - \frac{{{\hbar
^2}}}{{2m}}{\nabla ^2} + V({\bf{r}}) + U{{\left| {\psi
({\bf{r}})} \right|}^2}} \right]\psi ({\bf{r}}) = \mu \psi(\bf{r}), 
\end{eqnarray}
where $m$ is the particle mass, $V$ is the trapping potential,
$U=4 \pi \hbar^2 a_s /m$ is the interaction strength
(proportional to the two-body scattering length $a_s$), and $\mu$ is the chemical potential of the equilibrium system \cite{pethick_smith_2008}. We further employ dimensionless units rescaled by an oscillator length $S_l
= \sqrt{\hbar/(2m\omega_0)}$, where $\omega_0$ is a relevant frequency.

In order to access both the harmonically trapped filled-sphere condensate as well as hollow condensates of arbitrary thickness and radius, we use a spherically symmetric bubble trap~\cite{Zobay2001} for the trapping potential:
\begin{equation}\label{eq:bubble}
V(r)=m\omega_0^2 S_l^2
\sqrt{(r^2-\Delta)^2/4+\Omega_b^2},
\end{equation}
where $\Delta$ and $\Omega_b$ are the effective (dimensionless)
detuning between the applied rf field and the energy states used
to prepare the condensate and the Rabi coupling between these
states, respectively. The minimum of this potential is
found at $r=\sqrt{\Delta}$ and the frequency of single-particle
small oscillations around this minimum is $\sqrt{\Delta/\Omega_b}
\omega_0$. 

To fix a vortex of angular momentum $\ell$ along the $z$-axis in the system (and noting that the potential is spherically symmetric), we take a condensate wave function of the form
\begin{equation}
\psi({\bf{r}})=f(r)e^{i\ell \phi}
\end{equation}
where $\phi$ is the usual azimuthal coordinate. Plugging this form into Eq.~(\ref{eq:3DGPeqn}) above results in an equation for the condensate amplitude $f$ and energy $\mu = E_\ell$. For any specific value of $\ell$, the energy and amplitude can be straightforwardly found numerically using the imaginary-time algorithm described in Ref.~\cite{Chiofalo}. In this work, we have used a relatively high dimensionless interaction strength $u=8\pi N a_s/S_l=10,000$ and taken $\Delta/\Omega_b=1$ for simplicity. For a filled sphere, we take the bubble-trap parameter $\Delta=0$, while the thin shell results reported here were obtained for $\Delta = 200$.

\bibliography{reference}

\begin{thebibliography}{61}
\bibitem [{2D_()}]{2D_numerical_method}%
  \BibitemOpen
  \href@noop {} {}\bibinfo {note} {The key step of energy minimization is
  performed with imaginary-time evolving a trial wave function as
  $\psi_{2D}(\tau+\delta\tau)= e^{-\delta\tau H}\psi_{2D}(\tau)$, where $H$ is
  the Hamiltonian producing the energy functional of
  Eq.~(\ref{eq:2D_energy_functional}).}\BibitemShut {Stop}%
\end{thebibliography}

\end{document}